# 3 kV AlN Schottky Barrier Diodes on Bulk AlN Substrates by MOCVD

Dinusha Herath Mudiyanselage, Dawei Wang, Ziyi He, Bingcheng Da, and Houqiang Fu, *Member, IEEE*

*Abstract*—This letter reports the first demonstration of AlN Schottky barrier diodes on bulk AlN substrates by metalorganic chemical vapor phase deposition (MOCVD) with breakdown voltages exceeding 3 kV. The devices exhibited good rectifying characteristics with ON/OFF ratios on the order of $10^6$ to $10^8$ and excellent thermal stability from 298 to 623 K. The device Schottky barrier height increased from 0.89 to 1.85 eV, and the ideality factor decreased from 4.29 to 1.95 with increasing temperature, which was ascribed to the inhomogeneous metal/AlN interface. At reverse bias of –3 kV, the devices showed a low leakage current of 200 nA without the incorporation of any field plate structures or passivation techniques. This work demonstrates the potential of AlN as an ultra-wide bandgap semiconductor and represents a big step towards the development of multi-kV AlN high-voltage and high-power devices.

## I. INTRODUCTION

ALUMINUM Nitride (AlN) is a promising ultra-wide bandgap (UWBG) semiconductor for next-generation power electronics due to its remarkable material attributes, including the largest bandgap of 6.2 eV in the UWBG semiconductor family, high breakdown field of ~12–15 MV/cm, and superior thermal conductivity of 340 W/m·K [1]–[5]. In recent years, AlN Schottky barrier diodes (SBDs) have already shown decent progress with 1 kV breakdown voltages (BVs) and high-temperature stability [6]–[11]. Irokawa *et al*. [6] demonstrated lateral AlN SBDs on unintentional *n*-doped bulk AlN substrates with high-temperature stability up to 573 K. However, these devices showed a large ideality factor of 11.7, indicating the electron transport mechanism deviating from the well-known thermionic emission (TE) model. Kinoshita *et al*. developed [7] vertical AlN SBDs by removing the seed layer used to grow hydride vapor phase epitaxy (HVPE) AlN epilayers. These diodes were able to achieve BV between 550 – 770 V with an improved ideality factor of ~8. However, due to the lack of conductive AlN substrates, the substrate removal process used to fabricate vertical AlN SBDs could induce device damage. Fu *et al*. [8] reported lateral AlN SBDs with over 1 kV BV, showcasing the potential of AlN-based power electronics. Maeda *et al*. [9] recently showed molecular beam epitaxy (MBE) grown AlN SBDs with an AlGaN current spreading layer, and the devices exhibited good thermal stability and inhomogeneous apparent Schottky barrier. It was found that both forward and reverse current transports were dominated by defects in the AlN epilayers. The key to improving AlN device performance is to reduce defects in the material. Recently, HexaTech Inc and Crytal IS have commercialized physical vapor transport (PVT) grown bulk AlN substrates with low dislocation densities on the order of $10^3$–$10^4$ cm$^{-2}$ [9], [12]. This enables the growth of high-quality AlN epilayers by MBE and metalorganic chemical vapor phase deposition (MOCVD), where the latter is the industrial standard tool for mass production. In this work, we demonstrated the first 3 kV AlN SBDs on bulk AlN substrates by MOCVD with excellent high-temperature performance. The devices showed good rectifying behaviors with ON/OFF ratios of $10^6$–$10^8$ from 298 to 623 K and good thermal stability. At –3 kV, the devices exhibited a low leakage current of 200 nA without using any field plate structure or passivation. This work can serve as a reference for the development of multi-kV AlN high voltage high power devices.

## II. GROWTH, MATERIAL CHARACTERIZATION AND DEVICE FABRICATION

AlN epilayers were grown using MOCVD on (0001) bulk PVT AlN substrates. Trimethylaluminum (TMAl) and Ammonia (NH$_3$) were used as the Al and N sources respectively, whereas N$_2$ diluted silane (SiH$_4$) was used as the *n*-type dopant Si [13], [14]. The device structure, as illustrated in Fig. 1(a), consisted of a 1-µm-thick unintentionally doped

This work is supported as part of ULTRA, an Energy Frontier Research Center funded by the U.S. Department of Energy, Office of Science, Basic Energy Sciences under Award # DE-SC0021230. This work is also partially supported by the National Science Foundation (NSF) under Award # 2302696.

Dinusha Herath Mudiyanselage, Dawei Wang, Ziyi He, Bingcheng Da, and Houqiang Fu are with the School of Electrical, Computer, and Energy Engineering, Arizona State University, Tempe, AZ 85287, USA (e-mail: houqiang@asu.edu).

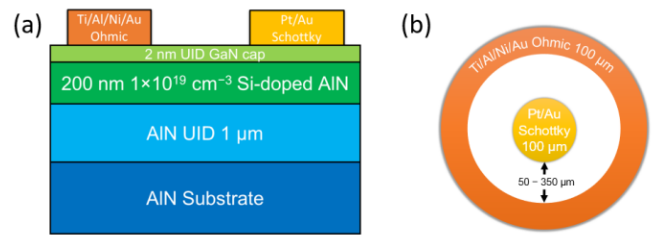

Fig. 1. (a) Schematic of fabricated AlN SBDs on bulk AlN by MOCVD with Ti/Al/Ni/Au Ohmic contact and Pt/Au Schottky contact. (b) Top view of devices.

(UID) AlN layer as a resistive buffer, a 200 nm highly Si-doped *n*-AlN layer, and a 2 nm UID GaN capping layer. The GaN capping layer was used to prevent oxidation of the underlying AlN epilayers upon exposure to air, which could degrade device performance [15]. The Si doping concentration in the *n*-AlN layer was $1\times10^{19}$ cm$^{-3}$. To study the crystal quality of the MOCVD-grown AlN sample, high-resolution X-ray diffraction (HRXRD) measurements were conducted using the Rigaku SmartLab X-ray diffractometer system. Figures 2(a) and 2(b) depict the (0002) symmetric and ($10\bar{1}2$) asymmetric rocking curves (RCs) for the AlN sample, with a full-width half

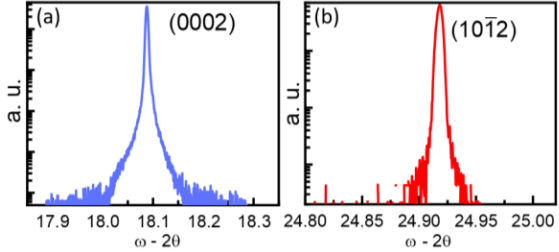

Fig. 2 (a) Rocking curve of (0002) plane and (b) ($10\bar{1}2$) for AlN epilayers by HRXRD.

maximum (FWHM) of 17.6 arcsec for (0002) and 19.08 arcsec for ($10\bar{1}2$). The dislocation density was estimated to be in the range of $10^4$–$10^5$ cm$^{-2}$ using the equations in Ref. 16. Furthermore, the surface morphology of the AlN sample was assessed using Bruker's Dimension atomic force microscopy (AFM), revealing a root-mean-square (RMS) roughness of ~ 1.2 nm over a 2×2 μm$^2$ scanning area. These HRXRD and AFM results indicate that the MOCVD-grown AlN epilayers possessed a low dislocation density and a smooth surface. The as-grown sample underwent a cleaning process involving acetone, isopropyl alcohol, and deionized water aided by ultrasonication. Subsequently, it was immersed in a hydrochloric acid (HCl) solution with a 1:2 (HCl:H$_2$O) ratio. The fabrication of AlN SBDs was performed using conventional optical photolithography and lift-off processes. Ohmic contacts were formed using Ti/Al/Ti/Au (25/100/25/50 nm) metal stacks deposited via electron beam deposition, followed by rapid thermal annealing (RTA) at 1000 °C in N$_2$ for 1 minute. The circular Ohmic contact had a width of 100 μm [Fig. 1(b)]. For Schottky contacts, Pt/Au (30/120 nm) metal stacks were deposited via electron beam evaporation. The distance between the anode and cathode contacts *d*, was varied between 50–350 μm. No field plate, passivation, or edge termination structures were implemented on the devices. Electrical measurements were performed on a probe station equipped with a Keithly 4200 SCS semiconductor analyzer and a thermal chuck. Reverse I–V characteristics were measured using Keysight B1505A Power Device Analyzer/Curve Tracer, and reverse breakdown measurements were conducted in insulating Fluorinert liquid FC-70 at room temperature.

## III. RESULTS AND DISCUSSIONS

Figure 3(a) shows the forward I–V characteristics of the AlN SBDs. The devices showed ON/OFF ratios on the order of $10^5$–$10^6$ and turn-on voltage of ~2.5 V, which are comparable to those of previously reported AlN SBDs [6]–[9]. The general diode equation for an SBD [17] can be written as

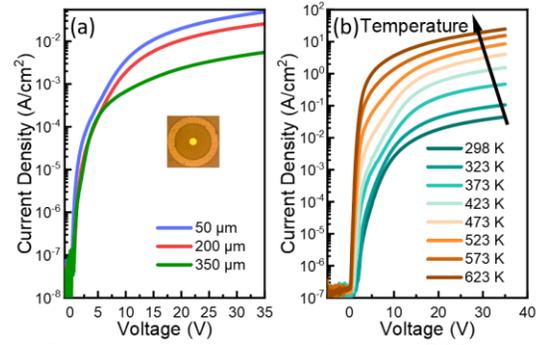

Fig. 3 (a) Forward I–V characteristics of the AlN SBDs with different contact distances on a log scale. Inset: An optical image of fabricated AlN device. (b) Temperature-dependent forward I–V characteristic of the AlN SBDs with a contact distance of 50 μm.

$$J = J_s \left[\exp\left(\frac{qV}{nkT}\right) - 1\right] \quad (1)$$

$$J_s = A^* T^2 \exp\left(-\frac{q\varphi_b}{kT}\right) \quad (2)$$

where $J$ is the current density, $J_s$ is the saturation current density, $A^*$ is the Richardson constant, $T$ is the temperature in Kelvin, $q$ is the electron charge, $\varphi_b$ is the Schottky barrier height, $n$ is the ideality factor, $k$ is the Boltzmann constant, $m^*$ is the effective electron mass, and $h$ is the Planck constant. The Richardson constant used in the calculation was 57.7 Acm$^{-2}$K$^{-2}$ using the effective electron mass of $0.48m_0$ [7] where $m_0$ is the free electron mass. Based on Eqs. (1), and (2), similar Schottky barrier height ($\varphi_b$) of ~ 0.9 eV was obtained for the devices with contact distances of 50, 200, and 350 μm. However, the ideality factor ($n$) varied with increasing distance between Ohmic and Schottky contacts. The minimum value of 4.29 was obtained for the devices with $d$ = 50 μm, which is comparable to previously obtained $n$ for AlN devices [8], [9]. However, other devices with larger contact distances exhibited a slightly larger $n$ (6.11 and 7.52 for $d$ = 200 and 350 μm devices, respectively). This indicates that the current transport mechanism is likely to be influenced by surface states and/or resistance of the AlN epilayers due to relatively low carrier concentration. Figure 3(b) shows the I–V curves of AlN SBDs at different temperatures. The devices showed good temperature stability from 298 up to 623 K, and the device ON/OFF ratio increased from $10^6$ to $10^8$ as more carriers contribute to the current transport at higher temperatures.

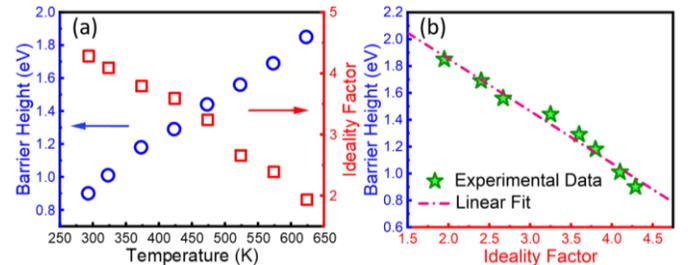

Fig. 4. (a) Ideality factor and Schottky barrier height of the AlN SBDs as a function of temperature. (b) Schottky barrier height vs ideality factor.

Figure 4(a) shows the temperature-dependent $\varphi_b$ and $n$ of the AlN SBD ($d$ = 50 μm). The $\varphi_b$ increased from 0.89 to 1.85 eV, and $n$ decreased from 4.29 to 1.95 with increasing temperature. Figure 4(b) shows a linear relationship between $\varphi_b$ and $n$ of the devices. This can be ascribed as an inhomogeneous metal/AlN interface with distributed low and high Schottky barrier regions. This behavior has also been commonly observed in previous reports [8], [9], [18]. As temperature increases, electrons can overcome higher Schottky barrier regions, leading to an increase in $\varphi_b$. In addition, the devices exhibited behaviors that are more closely aligned with the TE model with increasing temperature, as evidenced by the decreasing $n$ [9].

Furthermore, C–V measurements can be used to extract the carrier concentration of the AlN epilayers using the following equations [17]

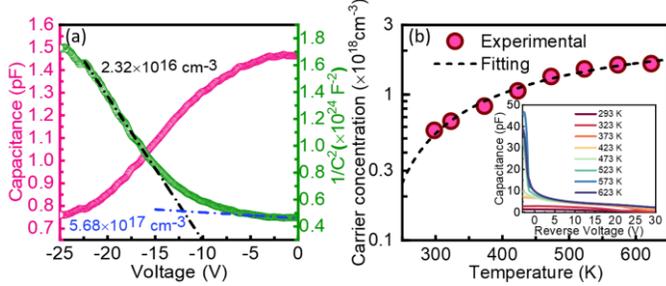

Fig. 5 (a) C–V plot and 1/$C^2$–V plot at RT for the AlN SBDs. (b) Carrier concertation vs. temperature plot. Inset: Temperature-dependent C-V measurement of the devices.

$$\frac{1}{C^2} = \frac{2}{q\varepsilon_0\varepsilon_r N_D}(V_{bi} - V - kT/q) \quad (3)$$

$$d\left(\frac{1}{C^2}\right)/dV = -\frac{2}{q\varepsilon_0\varepsilon_r N_D} \quad (4)$$

where $V_{bi}$ is the built-in voltage, $\varepsilon_0$ is the permittivity of the vacuum, and $\varepsilon_r$ is the relative permittivity of AlN ($\varepsilon_r$ = 9.2) [19]. Figure 5(a) shows the C–V and 1/$C^2$–V plots of the device. C–V measurements of the devices were performed at 10 kHz. The 1/$C^2$–V plot had two regions, corresponding to the AlN UID layer and $n$-doped AlN, respectively. The extracted carrier concentration of the UID layer was $2.3\times10^{16}$ cm$^{-3}$, whereas the $n$-doped region has a carrier concentration of $5.7\times10^{17}$ cm$^{-3}$, which is much smaller than the Si doping concentration due to dopant compensation in AlN [20], which leads to high Si donor ionization energy in AlN [14]. Figure 5(b) shows temperature–dependent carrier concentration extracted from temperature–dependent C–V measurements of the devices as shown in the inset. As the temperature increases the carrier concentration varied between $5.7\times10^{17}$–$1.6\times10^{18}$ cm$^{-3}$ from 298 to 623 K. As the temperature increases, more carriers are excited and contribute to the additional capacitance observed.

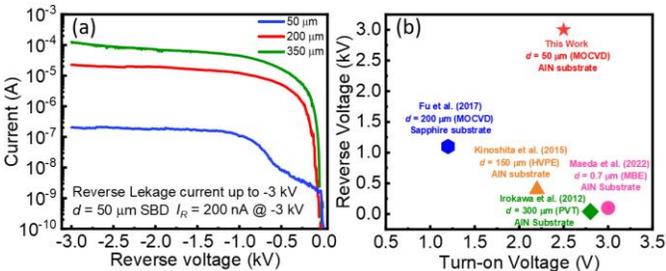

Fig. 6. (a) Reverse leakage of AlN SBDs. (b) Comparison of the breakdown and turn-on voltages of reported AlN SBDs with the pad distance and growth method.

Figure 6(a) shows the reverse I–V characteristics of the AlN SBDs with different contact distances up to −3kV. All the devices exhibited BV of over 3 kV. It should be noted that no destructive breakdown of the devices was observed up to −3 kV (the limit of the current setup). The AlN SBD with a contact distance of 50 μm showed a low reverse leakage of ∼ 200 nA at −3kV. The reverse leakage current increased with increasing contact distance, indicating that surface leakage is dominant and increases with the area of the devices. Figure 6(b) compares the BV and turn-on voltages of reported AlN SBDs. Our work showed record-high BV of over 3 kV with the smallest contact distance and comparable turn-on voltages.

## IV. CONCLUSION

In summary, lateral 3 kV AlN SBDs were grown and fabricated on bulk AlN substrates by MOCVD. The devices showed excellent rectifying behaviors with ON/OFF ratios of $10^6$–$10^8$ from 298 to 623 K and good thermal stability. With increasing temperature, $\varphi_b$ increased from 0.89 to 1.85 eV, while $n$ decreased from 4.29 to 1.95. At −3 kV, devices exhibited a low leakage current of 200 nA without any field plate structure or passivation. These results show the great potential of AlN and serve as an important reference for the future development of multi-kV AlN power electronics.